\documentclass[aps,pre,preprint,groupedaddress]{revtex4}
\usepackage{graphicx}
\usepackage{subfigure}

\begin{document}
\title{Effects of stochastic population fluctuations in two models of biological macroevolution}

\author{Y.~Murase}
\email[]{murase@serow.t.u-tokyo.ac.jp}
\author{T.~Shimada}
\email[]{shimada@ap.t.u-tokyo.ac.jp}
\author{N.~Ito}
\email[]{ito@ap.t.u-tokyo.ac.jp}
\affiliation{Department of Applied Physics, School of Engineering, The University of Tokyo,
7-3-1 Hongo, Bunkyo-ku, Tokyo 113-8656, Japan}
\author{P.~A.~Rikvold}
\email[]{rikvold@scs.fsu.edu}
\affiliation{
School of Computational Science, Center for Materials Research and Technology, and Department of Physics, Florida State University, 
Tallahassee, FL 32306-4120, USA}

\date{\today}

\begin{abstract}
Two mathematical models of macroevolution are studied.
These models have population dynamics at the species level, 
and mutations and extinction of species are also included.
The population dynamics are updated 
by difference equations with stochastic noise terms 
that characterize population fluctuations.
The effects of the stochastic population fluctuations 
on diversity and total population sizes 
on evolutionary time scales are studied.
In one model, species can make either predator-prey, mutualistic, 
or competitive interactions,
while the other model allows only predator-prey interactions.
When the noise in the population dynamics is strong enough, 
both models show intermittent behavior and their power spectral densities show 
approximate $1/f$ fluctuations.
In the noiseless limit, the two models have different power spectral densities.
For the predator-prey model, $1/f^2$ fluctuations appears,
indicating random-walk like behavior, 
while the other model still shows $1/f$ noise.
These results indicate that stochastic population fluctuations 
may significantly affect long-time evolutionary dynamics.
\end{abstract}

\maketitle      

\section{Introduction}
	Biological macroevolution has attracted much interest,
	not only among biologists, but also among physicists, 
	because it is a highly nonlinear, far-from-equilibrium, 
	complex interacting system.
%
%
    	Recently, one of the authors 
	proposed a series of individual-based biological coevolution models 
	and studied their long-term statistical properties 
	\cite{rikvold2003pea,zia-jpa,0305-4470-38-43-005,rikvold:2007lr,rikvold2007ibp}.
	It was found that some properties are 
	observed universally for various models.
	One such property is an approximate $1/f$ power spectral 
	density (PSD) for the diversities (i.e., number of species).
	The models display intermittency in their time series: 
	relatively quiet periods are interrupted by active periods during which the community is rearranged. 
	During 
	the quiet periods, the system is considered to be in a 
	quasi-steady state (QSS).
	The $1/f$ fluctuations are so robust that one is reminded of 
	the concept of a universality class in critical phenomena.
	A naturally emerging question is whether this $1/f$ mode 
	is unique in evolving systems.
        In this short article, we report the existence of another mode.
	In that class, diversities and total population sizes show 
	$1/f^2$ PSDs, i.e., indicating random-walk like fluctuations.

\section{Models}

	The models we use in this article are modified versions 
	of individual-based coevolution models. 
	The original individual-based models have 
	a stochastic population dynamics 
	with discrete, non-overlapping generations.
	Each species is represented by a bit-string genome of length $L$.
	This $L$-bit genome supplies a pool of $2^L$ possible species.
	At the end of every generation, individuals of species $I$ 
	give rise to $F$ offspring before they die  
	with the reproduction probability $P_{I}$. 
	This depends on the population sizes 
	$n_J(t)$ of all the species present in the community at that time.
	With probability $(1-P_{I})$, the individual dies without offspring.
	The fecundity $F$ is taken as a constant, independent of $I$ and $t$ for simplicity. 
	In each generation, the genomes of the offspring
	mutate with probability $\mu/L$ per gene and individual. 
	Mutation is the origin of diversity, which is necessary for the evolutionary process.
	Thus all $n_J(t)$ are updated stochastically at the same time. 
	
	
	In this article, instead of individual-based population dynamics, 
	we consider a stochastic difference 
	equation to explicitly test the effect of stochasticity at the
	population level. 
	In the case of individual-based models, the 
	populations at $t+1$ follow a binomial distribution 
	with mean $n_I(t)P_I$ and variance $n_I(t)P_I(1-P_I)$. 
	We here approximate this process by replacing the binomial 
	distribution by a gaussian distribution.
	Populations of continuous variables are updated by the following stochastic difference equation:
	\begin{equation}
		n_I(t+1) = 
		F[ P_{I}n_{I}(t) + \kappa\sqrt{n_I(t)P_I(1-P_I)}
		\xi_I(t) ] \;,
	\end{equation}
	where $\kappa$ and $\xi_I(t)$ are parameters giving the noise level 
	and a random number drawn from 
	a gaussian distribution with mean $0$ and standard deviation
	$1$, respectively.
	The first and the second term in the equation represent the  
	mean and fluctuation of the number of individuals in generation
	$t+1$, respectively.
	When the noise coefficient $\kappa = 1$, 
	the model has the same mean and 
	standard deviation as the corresponding individual-based model. 
	The model is then a good approximation for an individual-based model. 
	Populations tend to go to their equilibrium values,
	but fluctuate around them.
	This fluctuation may be critical,
	especially for species with tiny populations.
	On the other hand, when $\kappa=0$, the population dynamics
	becomes deterministic; 
	the system converges to a fixed point and does not 
	fluctuate after it reaches its fixed point.
	In this limit, a species can survive as long as it has a
	positive fixed point, 
	even when its equilibrium population is quite low.
	
	Extinction of species is introduced by defining a threshold 
	value, $n_{\rm th}=0.5$.
	A species whose population becomes less than $n_{\rm th}$ is 
	considered to go extinct and is removed from the system.
	In each generation, mutations happen after the population
	updates with a probability of $\mu/L$ per gene.
	The number of individuals is obtained by rounding off $n_I(t)$.
	
	The reproduction probability is given by
	\begin{equation}
		P_I{ \left\{n_J(t)\right\} } = \frac{1}{1+\exp{[ - \Delta_I( \{n_J(t)\})]} }.
	\end{equation}
	When $\Delta_I$ is large, species can almost certainly give rise to 
	offspring; 
	while they tend to die without offspring when $\Delta_I$ is
	small.

	We use two forms of $\Delta_I$, called model A and model B.
	Model A was introduced in \cite{rikvold2003pea}.
	In this model, $\Delta_I$ has the form 
	\begin{equation}\label{eq:delta_i_a}
		\Delta_I( \{n_J(t)\} ) = \sum_J M_{IJ} n_J(t)/ N_{\rm tot}(t) 
		- N_{\rm tot}(t)/N_0,
	\end{equation}
	where $M_{IJ}$, $N_{\rm tot}$, and $N_0$ denote the interaction 
	coefficient between $I$ and $J$, 
	the total population $\sum_I n_I(t)$, and the environmental 
	carrying capacity, respectively.
	The last term in Eq.~(\ref{eq:delta_i_a}) 
	limits the total population to a finite value.
	All the off-diagonal elements of the interaction matrix $M_{IJ}$ 
	are randomly drawn from a uniform distribution over $[-1,+1]$, 
	while the diagonal elements $M_{II}$ are set to $0$.
	
	In model B, $\Delta_I$ is
	\begin{equation}\label{eq:delta_i_b}
		\Delta_I( R, \{n_J(t)\} ) = - b_I + \eta_I R / N_{\rm tot}(t) 
		+ \sum_J M_{IJ} n_J(t)/ N_{\rm tot}(t),
	\end{equation}
	where $b_I$ and $\eta_I$ denote the reproduction cost and 
	the ability to utilize the external resource for species $I$, respectively.
	The external resource is represented by $R$, which remains
	constant.
	The birth costs $b_I$ are randomly drawn from a uniform distribution
	over $[0,1]$.
	A certain proportion of species ($0.05$) has the 
	ability to utilize the external resource.
	Thus $\eta_I$ is positive only for these species, and it is then 
	drawn from a uniform distribution over $[0,1]$.
	Other species have $\eta_I=0$.
	The interaction matrix is limited to anti-symmetric form 
	$(M_{IJ} = -M_{JI})$, 
	and it is non-zero with a certain connectance probability (here, $0.1$).
	Each off-diagonal $M_{IJ}$ 
	is selected from a triangular distribution over $[-1,1]$.
	The diagonal elements are distributed uniformly over $[-1,0]$.
	See \cite{rikvold2007ibp} for further details.

\section{Results}
	We performed Monte Carlo simulations and calcualted 
	the PSDs of the diversity and the total population.
	We used the diversity measure known in ecology as the 
	Shannon-Wiener diversity,
	which is defined as the exponential of the information-theoretical entropy of the population distribution 
	$D(t) = \exp [S(\{n_I(t)\}]$, where 
	\begin{equation}
		S\left( \left\{ n_I\left(t\right) \right\} \right) 
		= -\sum_{ \left\{I|\rho_I(t)>0\right\} } \rho_I(t)\ln{ \rho_I(t) },
	\end{equation}
	with $\rho_I(t) = n_I(t)/N_{tot}(t)$.
	Since there are many kinds of species with tiny populations that are mostly unsuccessful mutants, 
	this measure of diversity is useful to filter out the
	corresponding noise.

\begin{figure}[htbp]
	\subfigure[PSD of diversities]{
	\includegraphics[width=.48\textwidth]{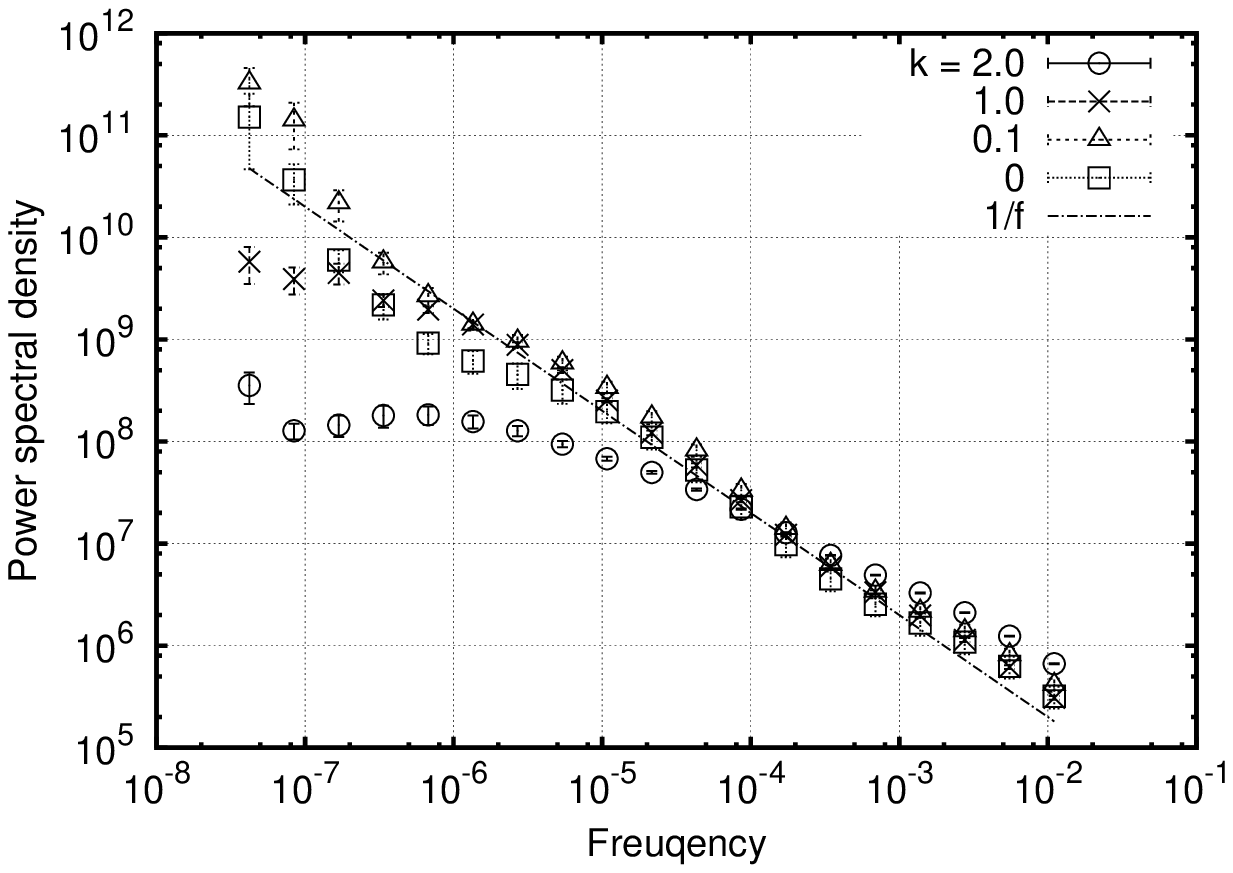}
	}
	\subfigure[PSD of total population sizes]{
	\includegraphics[width=.48\textwidth]{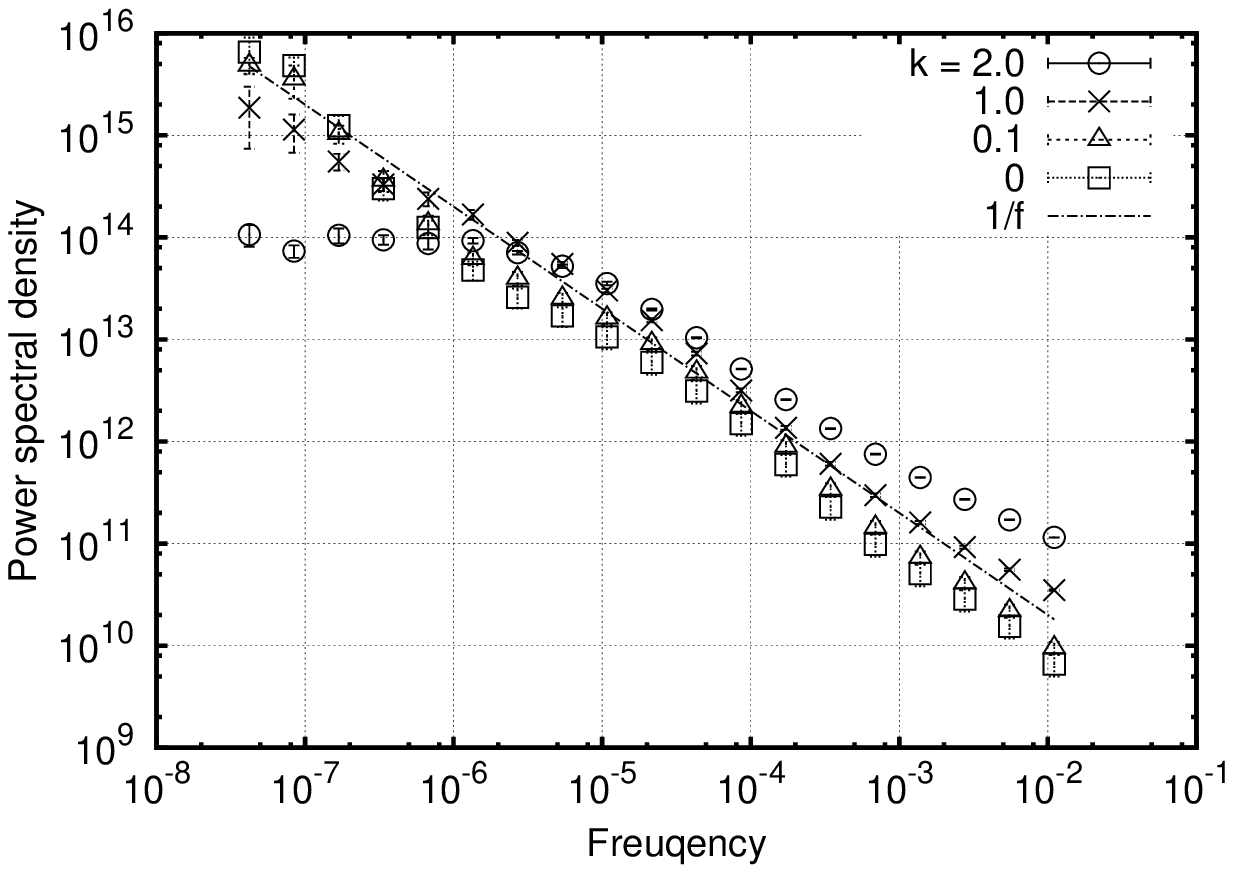}
	}
	\caption{
	Power spectral densities of diversities (a)
	and total population sizes (b) for model A at several noise levels.
	A line indicating $1/f$ is also shown in both figures as 
	a guide to the eye.
	}
	\label{fig:result_a}
	\end{figure}

	For model A, we performed 12 independent runs for each noise level.
	Each simulation run was performed for a long period of 
	$2^{25} = 33\,554\,432$ generations  
	with an initializing period of $2^{24}=16\,777\,216$ generations.
	The simulation parameters were 
	$L=13$, $F=4$, $N_0=2000$, and $\mu=0.001$.
	Results for several noise levels are shown in Fig.~\ref{fig:result_a}.
	It shows approximate $1/f$ behavior, regardless of the noise
	level. However, a crossover to  
	white noise is seen in the low-frequency range for strong noise,
	indicating the appearance of a characteristic time scale.
	Thus, the $1/f$ intermittency is robust against the change in 
	noise level in this model.
	\begin{figure}[b]
	\subfigure[PSD of diversities]{
	\includegraphics[width=.48\textwidth]{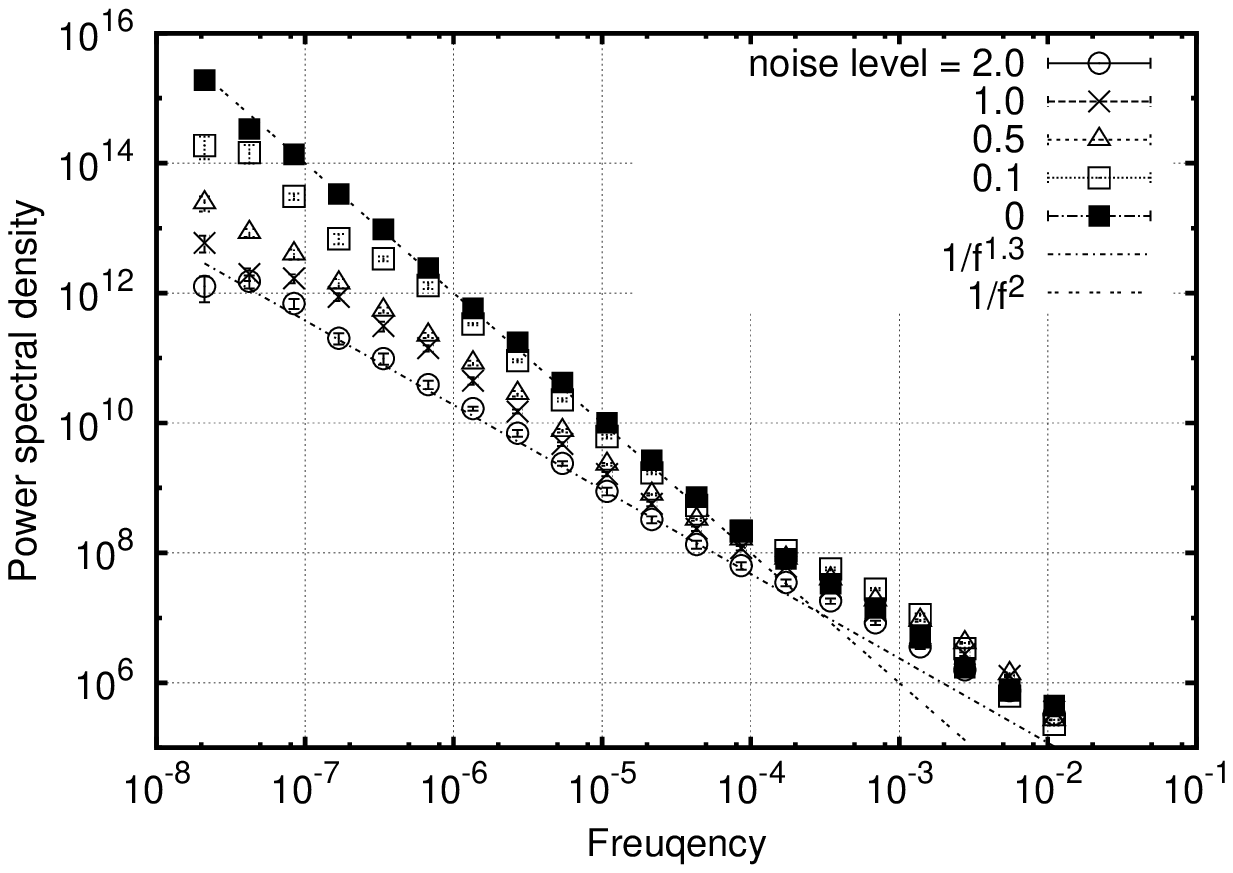}
	}
	\subfigure[PSD of total population sizes]{
	\includegraphics[width=.48\textwidth]{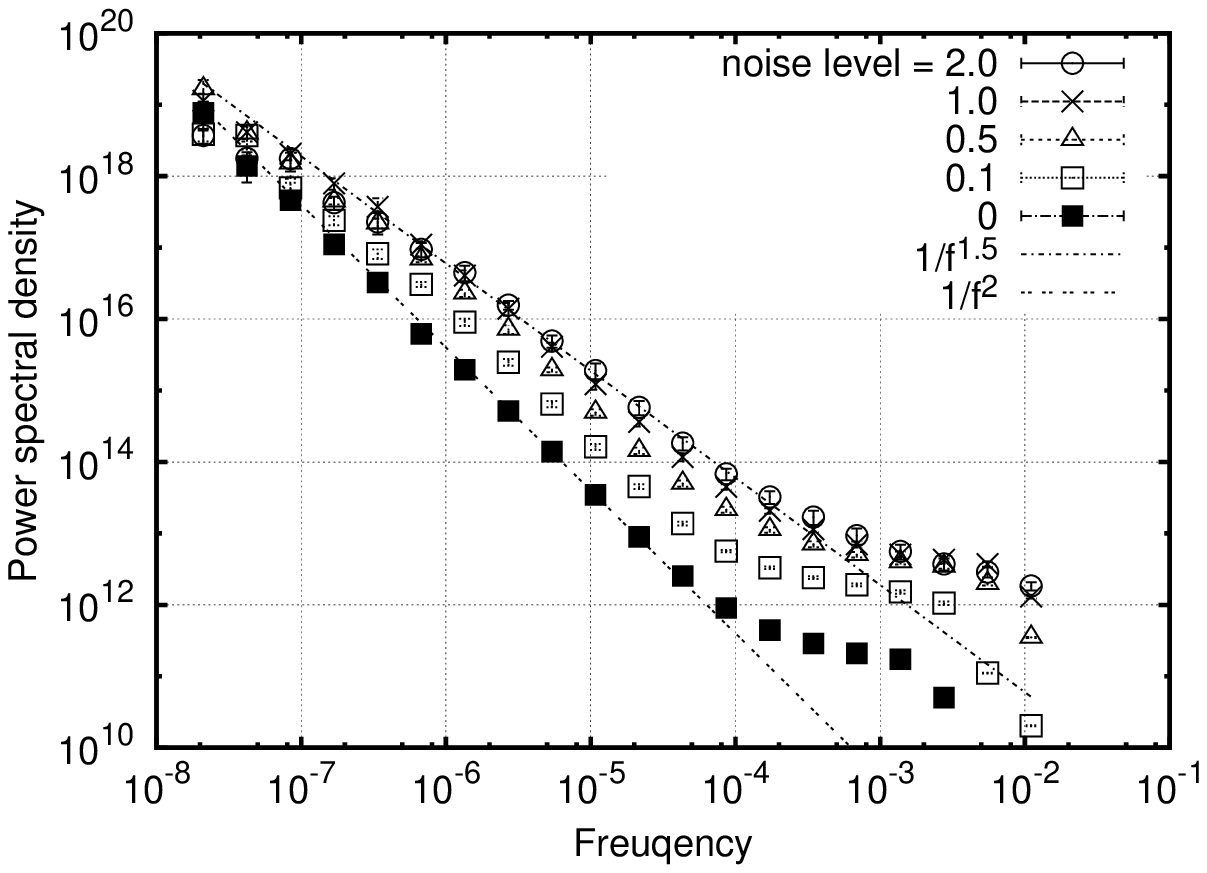}
	}
	\caption{
	Power spectral densities of diversities and total population sizes 
	with several noise levels for model B.
	Lines indicating $1/f^{\alpha}$ and $1/f^2$ are also shown in both 
	figures as guides to the eye,
	with $\alpha =1.3$ and $1.5$ for (a) and (b), respectively.
	}
	\label{fig:result_b}
	\end{figure}

	For model B, we performed 6 independent runs for each parameter.
	Simulations were performed for $2^{26} = 67\,108\,864$ generations 
	plus a $2^{24}$ generations initial ``warm-up" period.
	The parameters were $L=18$, $F=2$, $R=2000$, and $\mu = 0.0005$.
	The results are shown in Fig.~\ref{fig:result_b}.
	In contrast to model A, model B exhibits PSDs that depend on the
	noise level.
	When the noise is strong enough, they show approximate $1/f$ 
	fluctuations in a similar manner as the individual-based model.
	However, as the noise weakens, the exponents of the PSDs
	approach $2$; 
	and eventually they show $1/f^2$ like behavior in the noiseless limit. 
	The exponents change not suddenly but gradually.
	 A PSD with $1/f^2$ behavior means that the quantities change 
	 like random walks, 
	suggesting gradual changes of the species configurations.
	
%

\section{Summary and Discussion}
	In this short article, 
	we studied the effects of stochastic population fluctuations for two models of biological evolution.
	It was found that $1/f^2$ PSDs of diversity and total population sizes may appear in model B, 
	while $1/f$ PSDs are robustly found for model A.
	The evolving food webs in model B show not only $1/f$,
	but also $1/f^2$ behavior, depending on the level of the 
	stochastic population fluctuations.
	Therefore, the species configurations for model B in the noiseless 
	limit 
	change gradually rather than intermittently, 
	indicating the absence of QSS.
	It is desirable to analyze the noise sensitivity of model B in
	further detail to clarify the origin of the $1/f$ and $1/f^2$ fluctuations.
	For example, distributions of the lifetime of species or durations of QSS should provide further information.
	
	Other forms of population dynamics including more simplistic ones \cite{shimada-arob2002,shimada-arob2006} 
	and realistic ones \cite{drossel01:_influen_of_predat_prey_popul,Drossel:2004fj} have also been suggested.
	Comparison with these forms is a promising way to confirm the robustness of the results shown in this article.
	

%

\section*{Acknowledgments}
This work was supported by 21st Century COE Program ``Applied Physics on Strong Correlaiton''
from the Ministry of Education, Culture, Sports, Science, and Technology of Japan.
Work at Florida State University was partly supported by U.S.\ NSF Grant
No.\ 0444051.

%


\begin{thebibliography}{9.}
\addcontentsline{toc}{section}{References}

\bibitem{rikvold2003pea}
P.~A.~Rikvold and R.~K.~P.~Zia: Phys.~Rev.~E \textbf{68}, 031913 (2003)

\bibitem{zia-jpa} 
R.~K.~P.~Zia and P.~A.~Rikvold: J.~Phys.~A \textbf{37}, 5135 (2004)

\bibitem{0305-4470-38-43-005}
V.~Sevim and P.~A.~Rikvold: J.~Phys.~A \textbf{38}, 9475 (2005)

\bibitem{rikvold:2007lr}
P.~A.~Rikvold: J.~Math.~Biol. \textbf{55}, 653 (2007)

\bibitem{rikvold2007ibp}
P.~A.~Rikvold and V.~Sevim: Phys.~Rev.~E \textbf{75}, 51920 (2007)

\bibitem{shimada-arob2002}
T.~Shimada, S.~Yukawa, and N.~Ito: Artif.~Life~Robotics \textbf{6}, 78 (2002)

\bibitem{shimada-arob2006}
T.~Shimada, Y.~Murase, S.~Yukawa, and N.~Ito: Artif.~Life~Robotics\textbf{11}, 153 (2007)

\bibitem{drossel01:_influen_of_predat_prey_popul}
B.~Drossel, P.~G.~Higgs, and A.~J.~McKane: J.~Theor.~Biol. \textbf{208}, 91 (2001)

\bibitem{Drossel:2004fj}
B.~Drossel, A.~J.~McKane and C.~Quince: J.~Theor.~Biol. \textbf{229}, 539 (2004)

\end{thebibliography}
\end{document}